\documentclass[letterpaper,12pt]{article}

\usepackage{scicite}

\usepackage{graphicx}
\usepackage[latin1]{inputenc}
\usepackage{amssymb}
\usepackage{color}
\def\ev #1{\left\langle #1 \right\rangle}

\newcommand{\var}{\textrm{Var}}

\usepackage{times}

\newenvironment{sciabstract}{
\begin{quote} \bf}
{\end{quote}}

\newcounter{lastnote}

\title{The Spontaneous Emergence\\
of Social Influence in Online Systems}

\author{J.-P. Onnela$^{1,2,3,4\ast}$ and F. Reed-Tsochas$^{2,5}$}

\date{}

\begin{document} 

\maketitle 

\vspace{-10mm}

\begin{center}
\begin{small}
\noindent $^{1}$Department of Physics, University of Oxford, Oxford OX1 3PU, UK\\
$^{2}$CABDyN Complexity Centre, Said Business School, University of Oxford, Oxford OX1 1HP, UK\\
$^{3}$Department of Biomedical Engineering and Computational Science,\\
Helsinki University of Technology, P.O. Box 9203, FIN-02015 HUT, Finland\\
$^{4}$Harvard Kennedy School, Harvard University, Cambridge, MA 02138, USA\\
$^{5}$Institute for Science, Innovation and Society, Said Business School,\\
University of Oxford, Oxford OX1 1HP, UK\\
$^\ast$To whom correspondence should be addressed; E-mail: Onnela@hcp.med.harvard.edu
\end{small}
\end{center}

\baselineskip24pt

\begin{sciabstract}
Social influence\cite{mason} drives both offline and online human behaviour. It pervades cultural markets, and manifests itself in the adoption of scientific and technical innovations as well as the spread of social practices\cite{granov}. Prior empirical work on the diffusion of innovations in spatial regions or social networks has largely focused on the spread of one particular technology\cite{coleman,griliches} among a subset of all potential adopters\cite{jerker}. It has also been difficult to determine whether the observed collective behaviour is driven by natural influence processes, or whether it follows external signals such as media or marketing campaigns\cite{bulte}. Here, we choose an online context that allows us to study social influence processes by tracking the popularity of a complete set of applications installed by the user population of a social networking site, thus capturing the behaviour of all individuals who can influence each other in this context. By extending standard fluctuation scaling methods\cite{taylor,eisler}, we analyse the collective behaviour induced by 100 million application installations, and show that two distinct regimes of behaviour emerge in the system. Once applications cross a particular threshold of popularity, social influence processes induce highly correlated adoption behaviour among the users, which propels some of the applications to extraordinary levels of popularity. Below this threshold, the collective effect of social influence appears to vanish almost entirely in a manner that has not been observed in the offline world.  Our results demonstrate that even when external signals are absent, social influence can spontaneously assume an on-off nature in a digital environment. It remains to be seen whether a similar outcome could be observed in the offline world if equivalent experimental conditions could be replicated.
 \end{sciabstract}

Social influence captures the ways in which people affect each others' beliefs, feelings, and behaviors. It has traditionally been in the domain of social psychology with principal focus on micro-level processes among individuals\cite{mason}, but it also plays a prominent role across the social sciences, for example, in the study of contagion in sociology\cite{granov}, herding behavior in economics\cite{avery}, speculative bubbles in financial markets\cite{shiller}, voting behavior\cite{lazarsfeld}, and interpersonal health\cite{obesity}. Social influence plays an especially important role in cultural markets\cite{salganik}, for products such as books and music,  and generally pervades any arena of life where the attitudes and tastes of individuals are influenced by others.

It is often useful to distinguish between local and global sources of influence, which typically are identified with an individual's interpersonal environment and the mass media, respectively\cite{katz}. The overall social influence arises from a mixture of local and global influences, which themselves emerge from different signals. The fact that these two processes operate at very different scales poses considerable challenges for the empirical study of social influence. For the purposes of our study, we define (i) \emph{local signal} as information on the behavior of individuals who are friends or acquaintances of ego, the person whose behavior is being analyzed, and (ii) \emph{global signal} as information on the aggregate behavior of the population. Note that these definitions rely on the potentially observable behaviors of others as opposed to the non-observable ones, such as their intentions or feelings. This framework incorporating local and global signals is very generic and possible behaviors range from the consumption of cultural products to making lifestyle choices. 

The structures of social influence are most naturally addressed from the perspective of social network analysis\cite{wass}. The notion of local influence presupposes that individuals are embedded in a social network that channels and directs how behaviors spread. Examples of such behaviors include innovation adoption among physicians\cite{coleman}, as well as other empirical and theoretical studies of diffusion\cite{griliches,everett,young,dodds}. The notion of global influence, on the other hand, presupposes that individuals have information on the aggregate popularity of products and behaviors. While a given social network can be used as a proxy for communicating the behavioral signals, one should ideally have access to a network that accurately represents the potential communication channels for a given local signal, and these channels may vary between different behaviors. In addition, individuals are often selective as to what information they choose to disclose to their friends, resulting in the local signal being necessarily incomplete, biased, or misrepresented\cite{friendsense}. Similarly, while accurate population level statistics exist for popular items, it is much harder to find statistics for more marginal products and behaviors.

A novel opportunity to study human behavior in a setting that overcomes these methodological limitations is provided by certain online environments. These systems have the advantage of allowing access to complete sub-populations of agents. When combined with appropriate tools of analysis, they enable the direct study of collective macro-level social behavior in very large social systems without sampling. We study a complete online social system with well-defined local and global signals by harnessing data from Facebook\cite{facebook}, a hugely popular social networking site (SNS), which at the time of data collection had approximately 50 million active users worldwide. In addition to the current popular interest in social networks, scholars have recognized the potential of these and other social websites for research\cite{mayer,christakis, huberman,sornette,traud}, reflecting the current move to utilising rich large-scale datasets on human behavior and communication\cite{onnela,lazer}. Facebook users, in line with other SNSs, can construct a public or semi-public profile within a bounded system, articulate a list of other users, ``Facebook friends'', with whom they share a connection, and view and traverse their list of connections and those made by others within the system\cite{boyd}.

Facebook users can also install (and uninstall) applications (Fig.~1A) that enable them, for instance, to play poker and compare their taste in movies with their friends\cite{facebook}. Whenever a user adopts a new application, her friends are automatically notified by the system\cite{jpnote}, but the users can also see the applications of any of their friends simply by visiting their profile. Consequently, users with many Facebook friends are then, at least in principle, in a position to influence a larger number of other users. In addition, everyone has access at all times to an all-inclusive listing of applications ranked by their global popularity, which acts as an effective ``best seller'' list. Although applications are free of charge, popular applications have the advantage of being readily discoverable (low search cost), and are more likely to be of higher quality both with respect to reliability (exhaustively tested) and functionality (superior features). The applications provide recreational value and can be seen as cultural goods, and the different ways the users process the local and global signals in choosing applications reflect their personal preferences, i.e. the underlying heterogeneity of the population. 

\begin{figure}
\begin{center}
\includegraphics[width=1\linewidth]{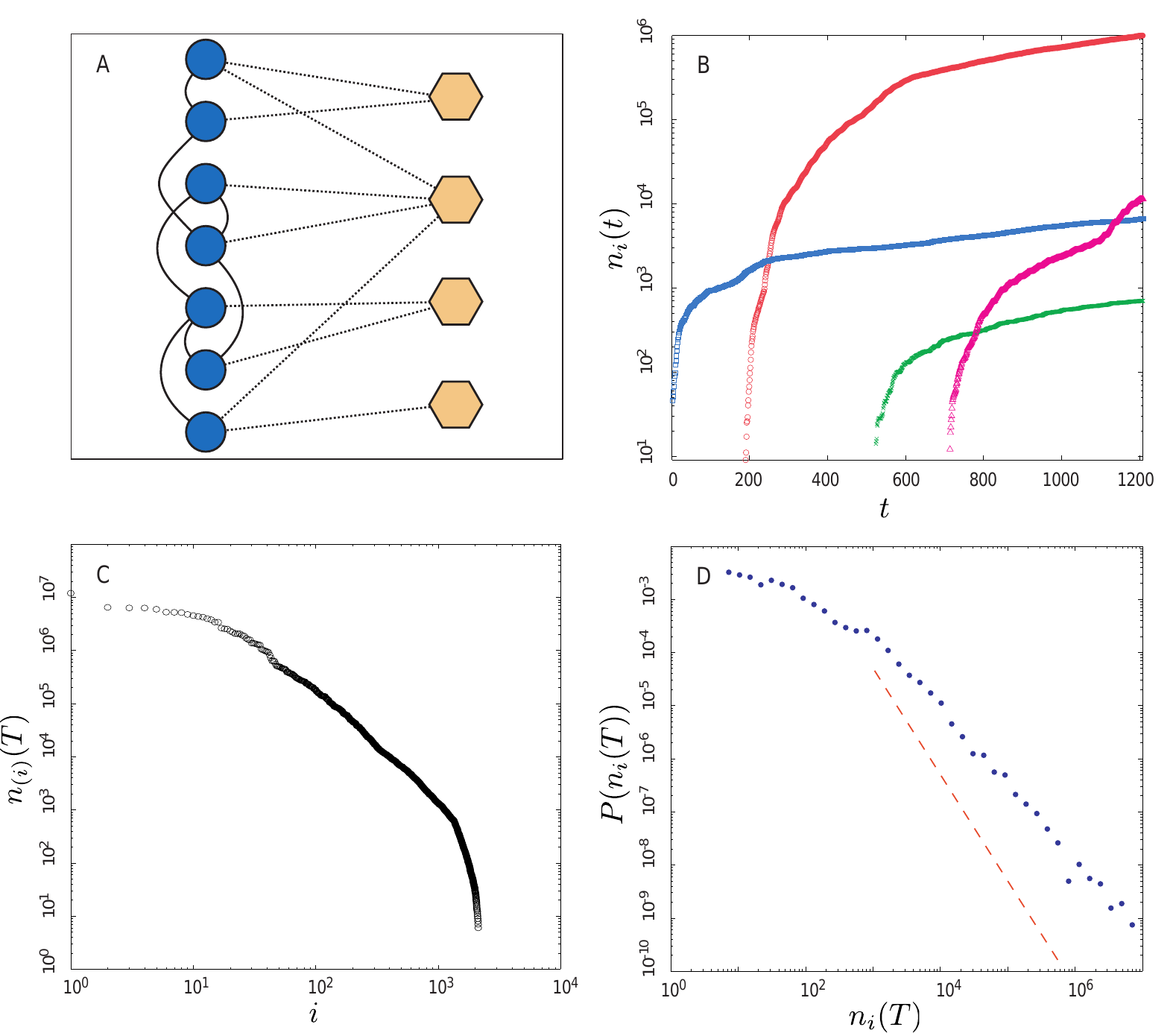}
\caption{
Facebook users and applications.
\textbf{(A)} The users (round nodes) form a social network (solid lines) which influences their behavior in adopting application (hexagons).
\textbf{(B)} Number of users $n_{i}(t)$ as a function of time $t$ for four applications of which ``Texas HoldEm Poker'' is the most popular one at the end.
\textbf{(C)} Number of users $n_{i}(T)$ sorted in descending order for the 2123 applications that have $n_{i}(T)>0 $ (Zipf plot). 
\textbf{(D)} Probability density distribution $P(n(T))$ vs. $n(T)$ is fat-tailed. The dashed line $\sim n(T)^{-2}$ is intended to guide the eye and corresponds to the limit where the mean of the distribution diverges.}
\end{center}
\end{figure}

In addition to the distinction between local and global signals, it is important to classify systems into two separate categories based on whether their dynamics are \emph{endogenous} without external drivers, or \emph{exogenous} and driven externally. Epidemic spreading in a closed system is an example of an endogenous process with local transmission, since the pathogens need to be passed from one person to another in close physical proximity. Similarly, it is possible to model the spread of innovations such as the uptake of new hybrid crops by farmers as an endogenous social contagion process, and to try to distinguish between different types of local processes that may underlie the observed rate at which the innovation is adopted\cite{young:diff}. However, studies of social influence which focus on local and endogenous processes such as word-of-mouth transmission are almost always open to the challenge that they neglect equally important exogenous effects such as marketing or mass advertising, and typically trying to separate these two confounding factors is highly problematic. For instance, a recent re-analysis\cite{bulte} of the classic diffusion studies on how prescriptions for an antibiotic drug spread among physicians in different communities \cite{coleman,griliches} suggests that marketing efforts, in this context corresponding to external drivers, can account for most of the observed behavior. In the current setup, both the local and global signals are generated endogenously within the system, i.e. there is no exogenous driver\cite{footnote}. 

We downloaded the data from Facebook for all existing 2720 applications between June 25, 2007 and August 14, 2007, shortly after applications were introduced. These data consist of time series $n_{i}(t)$ with $i=1,2,\ldots, M=2720$ and $t=1,2,\ldots,T=1208$ corresponding to the aggregate number of users who have application $i$ installed at time $t$ (Fig.~1B). Data for 15 applications were partly corrupted and were consequently omitted from the analysis, leaving us with 2705 applications, or $99\%$ of the data. Importantly, studying all the applications avoids a selection bias, which is generated by examining the trajectories of those applications that spread successfully as tends to be done in most studies on social influence\cite{jerker}. Successful products in  cultural markets have been found to be orders of magnitude more popular than the average cultural product \cite{salganik}. This finding is also manifest in the case of Facebook applications. The number of users at the end of the time horizon, $n_{i}(T)$, sorted in descending order is shown in Fig.~1C. For the ten most popular applications these numbers vary between $n_{(1)}(T) \approx 12$ million and $n_{(10)}(T) \approx 4.6$ million, whereas $n_{(100)}(T) \approx 180,000$ and $n_{(1000)}(T) \approx 1,300$. The probability density distribution for the number of application installations (Fig.~1D) has a very fat tail and decays so slowly that even its mean value diverges in the limit of infinite system size.

Each new installation, in addition to increasing the overall user base of the application and thus its global signal, also generates a local signal, through which the adopter may in turn influence the future behavior of his friends. Each installation thus acts as a microscopic social stimulus and creates a form of positive feedback in the system. Note that the observable behavior which generates patterns of social influence in this case is restricted to the adoption of an application, rather than its use. Given that the users are part of a very large social network, the consequences of adopting an application are not limited to a user's immediate neighborhood, but may percolate further in the network. This underlines the importance of having data that reflects the behavior of the entire network even if the underlying microscopic data are not available. While the impact of a single installation is admittedly minute, the superposition of the observed 104 million application installations leaves behind a detectable footprint.

To study the effect of social influence, i.e. the extent to which the behavior of an individual (his installing an application) is related to the behavior of others (their installing the same application), we turn to the method of fluctuation scaling (FS). This allows us to extract a key signature of the system's behavior purely on the basis of the above aggregate data. FS has been applied successfully to a number of complex systems whose interacting elements participate in some dynamic process. Examples of application domains range from fluctuations in population sizes in ecology to fluctuations in stock trading activity in financial markets\cite{smith,taylor,eisler}. Here we outline how FS can be utilized in the current problem, and refer the reader to Supplementary Information (SI) for details. For a given application $i$, the act of individual $j$ regarding installation of the application is encoded by the random variable $S_{i,j}(t)$, where $S_{i,j}(t) = {1}$ corresponds to him installing the application at time $t$, and $S_{i,j}(t) = {0}$ corresponds to him doing nothing. From the stochastic process point of view, one can think of each individual tossing coins at every time step, one per application, to decide whether he will install the given application. Social influence, operating through the local and global signals, is likely to render the coin tosses dependent for any given application (Fig.~2A,B). To measure the strength of social influence, we define \emph{net activity} $f_{i}(t)$ of application $i$ at time $t$ as 

\begin{equation}
f_i(t) \equiv n_{i}(t) - n_{i}(t-1) = \sum_{j=1}^{N} S_{i,j}(t) =  \sum_{k=1}^{N-n_{i}(t)} S_{i,j_{k}}(t),
\label{eq:f}
\end{equation}
which corresponds to the net increase in the number of installations for application $i$ between times $t-1$ and $t$. It can be expressed in terms of the individual constituent variables as shown, where the first sum is taken over all $N$ individuals, whereas the latter sum is taken over potential new installers, with the subset of indices $j_{1}, j_{2}, \ldots, j_{N-n_{i}(t)} \in \{1, 2, \ldots, N\}$ such that $S_{i,j_{k}}(t-1)=0$. In terms of the above analogy, once a user has installed a given application, he stops tossing the particular coin corresponding to that application.

\begin{figure}
\begin{center}
\includegraphics[width=1\linewidth]{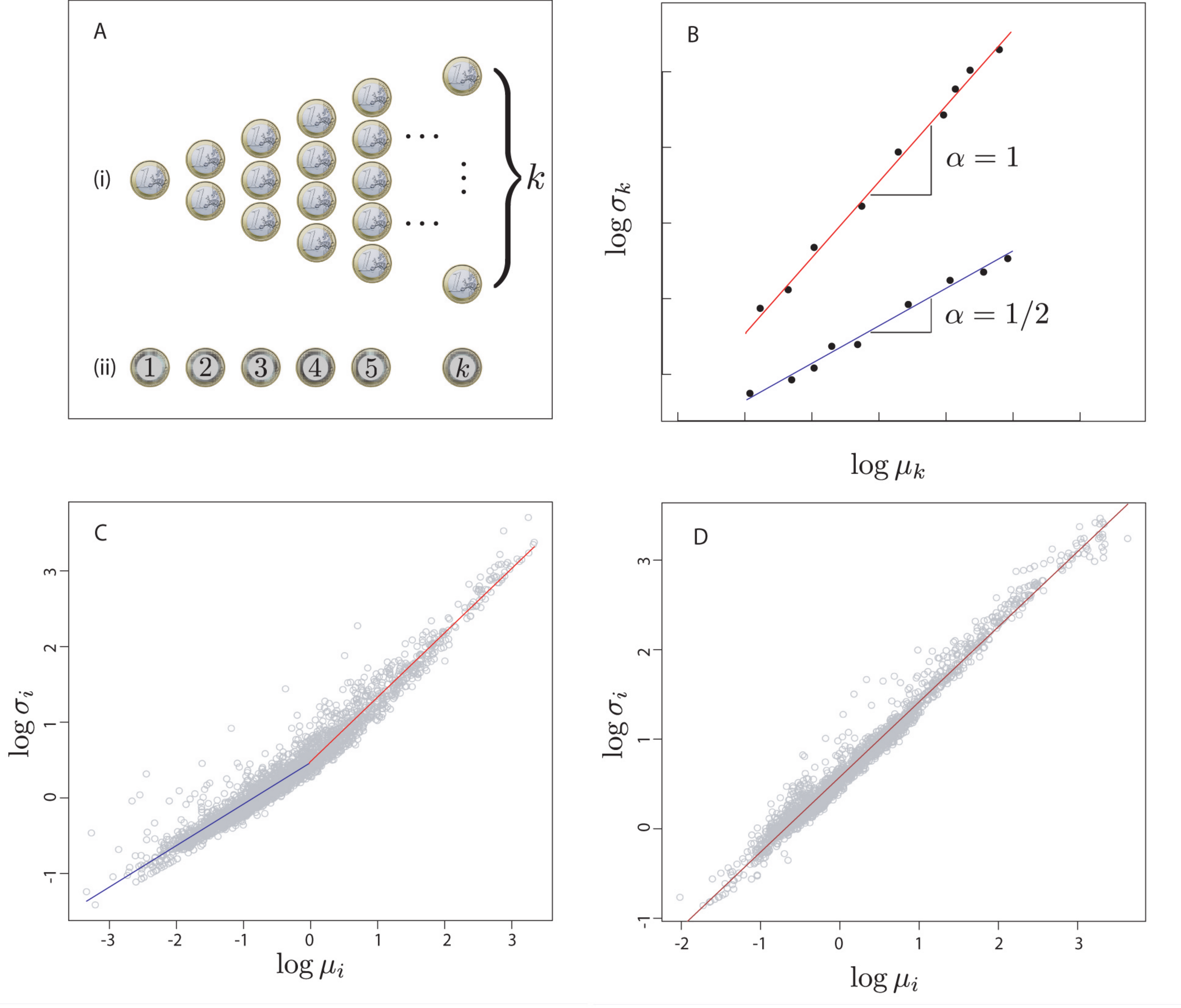}
\caption{Fluctuation scaling (FS). 
\textbf{(A)} The concept of FS can be illustrated by considering tossing coins in two ways \cite{eisler}. (i) We toss a group of $k$ coins independently with sides corresponding to 0 and 1 and let $f_{k}$ equal their sum. (ii) We toss a single coin with sides 0 and $k$, which corresponds to tossing $k$ fully coupled coins.
\textbf{(B)} We perform the experiment several times and calculate the average $\langle f_{k} \rangle$ and standard deviation $\sigma_{k}$ of $f_{k}$ as shown in the schematic. In both cases $\langle f_{k} \rangle \sim k$, whereas $\sigma_{k} \sim \sqrt{k}$ in (i) but $\sigma_{k} \sim k$ in (ii). Varying the value of $k$ produces a series of points in the $\log{\mu_{k}}$, $\log{\sigma_{k}}$ plane. From the FS point of view, this simple example resembles Facebook users making decisions on application adoption; the ``coins'' are now biased, reflecting individual heterogeneity, and the tosses are not independent but coupled via the local and global signals (see S2 in SOM).  
\textbf{(C)} Of the 2705 Facebook applications in the empirical data set, 2562 with $\mu_{i}>0$ and $\sigma_{i}>0$ are plotted here (see S3 in SOM). Two qualitatively different regimes emerge and they are separated by a cross-over point located at $\log\mu_{x} = 0.36$. The first, \emph{individual regime} is characterized by the exponent $\alpha_{I} \approx 0.55$, and the second, \emph{collective regime} by $\alpha_{C} \approx 0.85$.
\textbf{(D)} The synthetic data set consists of 2705 time series of which 2163 have $\mu_{i}>0$ and $\sigma_{i}>0$. We now obtain a single regime characterized by the exponent $\alpha_{S} \approx 0.84$. Note that in C and D the exponents lie between $1/2$ and $1$, corresponding to the extremes of completely uncorrelated and correlated decisions of users to adopt applications.}
\end{center}
\end{figure}

According to FS, the temporal average and standard deviation of $f_{i}(t)$ are related through the relationship $\sigma_{i} \sim \mu_{i}^{\alpha}$. This motivates us to identify a region in which the relationship between $\log{\mu_{k}}$ and $\log{\sigma_{k}}$ is linear. The value of the \emph{fluctuation scaling exponent} $\alpha$ is given by the slope of the line. Although $\alpha$ lies in the rather narrow range $[1/2,1]$, its value is crucial as an indicator of statistical coupling in the system (Fig.~2A,B). If the behavior of a user is independent of the behavior of others, one would expect $\alpha=1/2$, whereas if her behavior is fully correlated with others one would expect $\alpha=1$ for all applications\cite{footnote1}. 

As shown in Fig.~2C, applications with $\log(\mu_{i}) > \log(\mu_{x}) \approx 0.36$ define the \emph{collective regime} governed by $\alpha_{C} \approx 0.85$, which indicates strong correlations among the constituent variables, i.e. the underlying ``coin tosses''. Application installations above this point are influenced by the behavior of others. Unexpectedly and contrary to previous empirical studies of other systems\cite{eisler}, breakpoint analysis (see S4 in SI) shows that the system exhibits another qualitatively different regime for the less popular applications. This \emph{individual regime} with $\log(\mu_{i}) < \log(\mu_{x})$ has $\alpha_{I} \approx 0.55$, which is very close to the limiting case of $\alpha=1/2$, meaning that applications installations  are nearly uncorrelated and social influence is negligible. The transition between the two regimes takes place at approximately $\log(\mu_{x}) = 0.36$, which translates into an average daily activity of $24 \times 10^{0.36} \approx 55$ new installations a day. We emphasize that theoretical considerations guided our choice to fit a linear function to the data in Fig.~2C as opposed to, say, trying to find the best fit among a class of curvilinear functions. While it would be interesting to resolve also the precise location and nature of the transition (sharp or continuous), we are unable to make this distinction on the basis of the empirical data. However, the central finding on the existence of two different regimes remains unaffected.

The interpretation of FS exponents in terms of correlations assumes that the underlying stochastic processes is stationary\cite{eisler}. However, the fact that $n_{i}(t)$ increases over time demonstrates that the system cannot be stationary. The question then becomes whether the system is sufficiently close to stationarity so that the fluctuation scaling exponents can be given the above interpretation. Let us impose the stringent condition that the system is sufficiently close to stationarity when at most 1\% of users have the application installed. We show in SI that even under this strict condition, $98\%$ of the time series are stationary. This also means that the scaling in Fig.~2C holds for over two orders magnitude \emph{above} the cross-over point. We conclude that the system is sufficiently stationary so that the temporal fluctuations may indeed be given the above interpretation.

As a simple explanatory hypothesis for the observed behavior, one might suggest that the different scaling properties result from applications having different lifetimes. To test this, we divide the applications into three distinct groups based on their time of introduction such that each group covers an equally long time interval. We repeat the scaling plot by choosing randomly 300 applications from each group with the red, green, and blue colors indicating whether the application was introduced in the first, second, or third interval (Fig.~3). Since any interval of $x$-values contains an approximately equal number of markers of different colors, the time of introduction and, hence, application lifetime, does not explain its scaling properties.

\begin{figure}
\begin{center}
\includegraphics[width=0.8\linewidth]{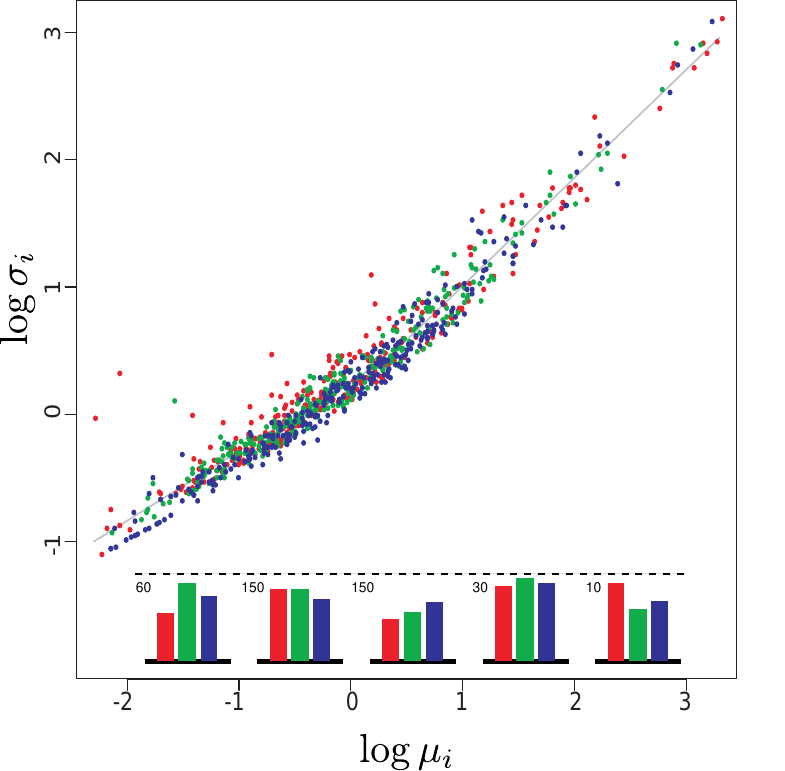}
\caption{Effect of application lifetime on scaling. Visual inspection shows that any interval of $\log \mu$-values contains a roughly equal number of red, green, and blue markers, indicating that the time of introduction and, hence, application lifetime, is not related to its scaling properties. The histograms at the bottom of the panel show exactly how many applications from each of the three periods (red, green, blue) fall in the $[-2,-1)$, $[-1,0)$, $[0,1)$, $[1,2)$, and $[2,3)$ intervals, demonstrating clearly that there is no age trend in the scaling plot.}
\end{center}
\end{figure}

Should the transition from one regime to the other be attributed to the popularity of applications reaching a certain threshold value, or should it be attributed to the system itself? If the former is true, then one might think that the transition corresponds to a phase transition or to the crossing of an epidemic threshold, essentially a density threshold effect, resulting in an epidemic of popularity. To isolate the effects of popularity, we construct rank-order preserving \emph{synthetic time series} from the empirical ones. This deterministic process (apart from ties) cuts the empirical time series into pieces and then recombines the pieces using a rank based rule (see Methods). As shown in Fig.~2D, the transition disappears for the synthetic data. Statistical tests also support the existence of a single regime (see SI) and, in addition, the Pearson's linear correlation coefficient between $\log{\sigma}$ and $\log{\mu}$ is $0.99$. The consequences of this are threefold. First, the lack of two regimes for the synthetic data demonstrates that the transition from one regime to the other is not a result of the popularity of an application exceeding a certain threshold, so the phenomenon is not analogous to crossing an epidemic threshold. Second, it demonstrates that the collective (correlated) regime does not result from the system becoming saturated with users of a given application that would then induce correlations between the behaviors of the individuals. This is because all the synthetic time series obey the same scaling relation also for small values of $\log(\mu)$ (corresponding to dilute limit), where the system is far from being saturated. Third, the synthetic regime has an exponent $\alpha_{S} \approx 0.84$, which is very close to $\alpha_{C} \approx 0.85$ that characterizes the collective regime for empirical data. This shows that we can recover the exponent of the collective regime by assuming that the future popularity of an application is driven by its current popularity, a finding that has also been used to predict popularity of online content\cite{gabor}.

We have harnessed data on Facebook applications to study the role of social influence on the dynamics of popularity in an endogenous online system. The way the platform, Facebook, and the cultural products, Facebook applications, have been set up in this self-contained system guarantees that the agents are subject to both local and global signals of influence. We have shown here that the studied online system exhibits a collective and individual regime, and argued that the emergence of the two regimes is an inherent property of the system. Since each regime is characterized by a single fluctuation scaling exponent, the strength of social influence is approximately constant across each regime. Consequently, the extent of social influence becomes discretized: either there is virtually no influence or, alternatively, the strength of influence is that given by the exponent of the collective regime. This suggests that social influence assumes a binary, on-off nature in the system. It is worth pointing out that had we only monitored the more successful (high $\mu$) applications, we would have been able to observe only (part of) the collective regime. 

We believe that our finding on the existence of the two regimes may well generalize to other systems. The move of an increasing number of human activities to the online world has endowed users with the power of participation. Familiar examples include the online book retailer Amazon and the online DVD rental service Netflix, both of which allow their users to rate the products and, consequently, influence their future popularity. While some books and films in these systems are certainly highly advertised by their producers, they arguably stand for only a small fraction of the choices available, leaving a large majority of books and films exposed to endogenously generated social influence. Social influence may then emerge spontaneously in a wide range of online environments over and above purely endogenous systems. Whether it becomes discretized in these systems as well remains to be seen.

\section*{Methods}

\subsection*{Synthetic time series}

\begin{figure}
\begin{center}
\includegraphics[width=1\linewidth]{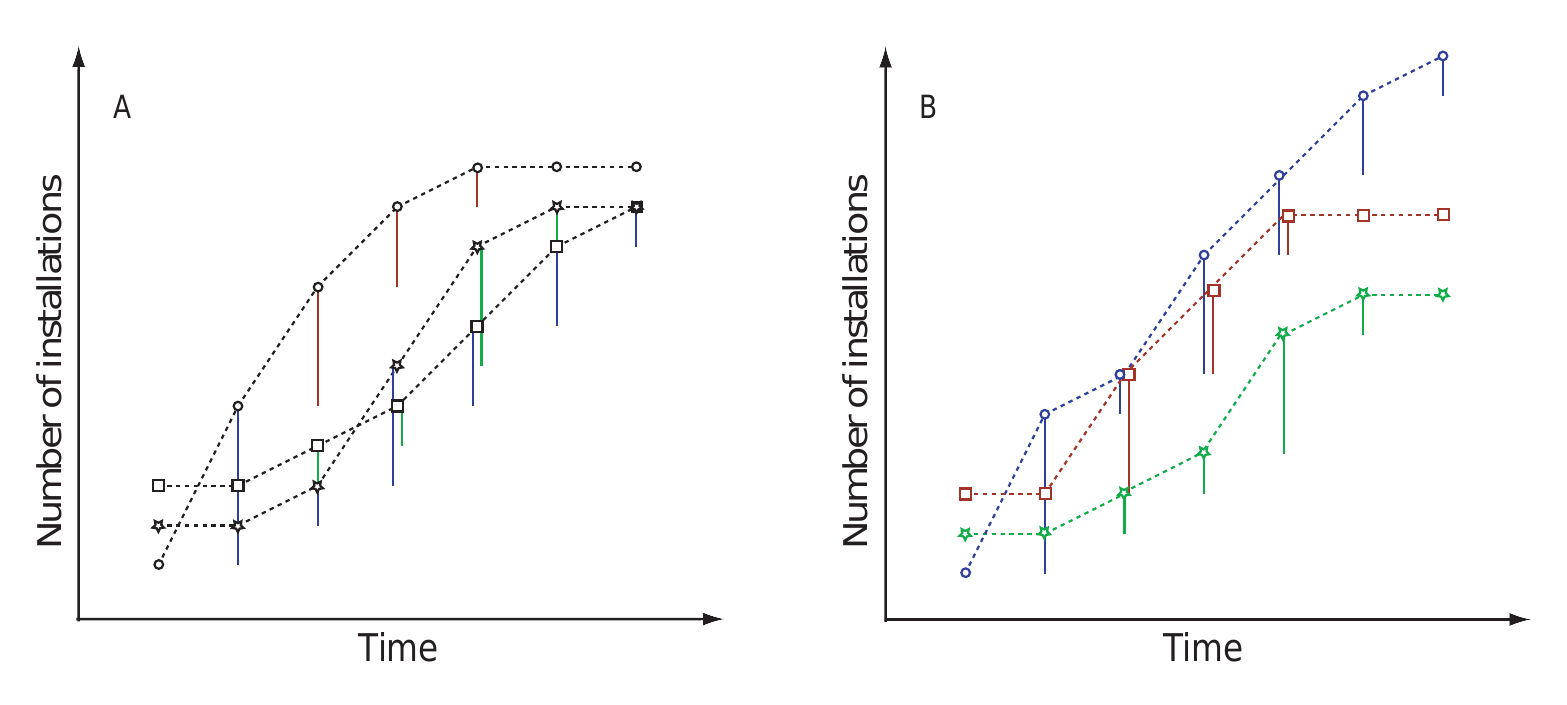}
\caption{Schematic of the construction of the synthetic time series $\tilde{n}_{i}(t)$. \textbf{(A)} The empirical data consists of $t=1,\ldots, 7$ observations for three applications. The data points have been connected with dashed black lines to guide the eye. For the most popular application at time $t-1$, the change in number of users  between times $t-1$ and $t$ is indicated by the height of the vertical red bar at time $t$, which corresponds to $\tilde{f}_{1}(t)$ in the text. Similarly, $\tilde{f}_{2}(t)$ and $\tilde{f}_{3}(t)$ are indicated by the green and blue bars, respectively. An easy way to understand the process is first to compute the difference in the number of users for all applications given by $f_{i}(t)=n_{i}(t)-n_{i}(t-1)$ and then color the difference based on $r_{i}(t-1)$, the rank of the application at time $t-1$. \textbf{(B)} The synthetic time series are seeded by the initial values taken from the empirical data such that $\tilde{n}_{1}(1) = n_{\Box}(1)$, $\tilde{n}_{2}(1) = n_{\star}(1)$, and $\tilde{n}_{3}(1) = n_{\circ}(1)$ of the empirical data and they are constructed by adding together the difference bars of the same color. Overlapping bars have been shifted slightly horizontally for clarity of presentation.
\label{fig:synth}}
\end{center}
\end{figure}

We construct rank-order preserving synthetic time series from the empirical time series in order to isolate the effects of popularity from other factors in the $\log{\sigma}$, $\log{\mu}$ plots. This process is deterministic (apart from ties) and essentially it cuts the empirical time series into pieces and then recombines the pieces using a rank based rule (see Fig. \ref{fig:synth}). Let us denote the global \emph{ranking} of application $k$ at time $t$ with $r_{k}(t) \in {1,\ldots,M}$ such that $n_{(k-1)}(t) \ge n_{(k)}(t) \ge n_{(k+1)}(t)$. We define $\tilde{n}_{i}(t) = \tilde{n}_{i}(t-1) + \tilde{f}_{i}(t)$ analogously to what we had before, but now $\tilde{f}_{i}(t) = n_{k}(t)-n_{k}(t-1)$ such that $r_{k}(t-1)=i$. Here $\tilde{f}_{i}(t)$ is the number of new installations over a single time step for an application that at time $t-1$ had ranking $i$. The series are seeded by setting $\tilde{n}_{i}(1) = n_{(i)}(1)$ for all $i=1,\ldots,M$ and are constructed using the above recursive relation for $t \ge 2$.

The synthetic time series $\tilde{n}_{i}(t)$ by construction has a constant relative popularity as measured by the global rank order of $\tilde{n}_{i}(t)$ and, consequently, the future popularity of the synthetic time series is systematically driven only by its current popularity (rank). In the absence of rank crossings, the synthetic data would behave like empirical data. The increments $\tilde{f}_{i}(t)$ of the synthetic data result from a combined effect of both the local and global signals. The impact of the global signal remains constant since the synthetic time series $\tilde{n}_{i}(t)$ always holds rank $i$ on the global ``best seller'' list. A single synthetic time series $\tilde{n}_{i}(t)$ is typically a combination of several empirical time series and, therefore, the local signal in the synthetic time series corresponds to a mean-field approximation of the local signals of the applications that make up the synthetic time series $\tilde{n}_{i}(t)$.

\bibliographystyle{Nature}

We are grateful to S. Gourley for the help with data acquisition, to D. Barron, J. Denrell, Z. Eisler, J. Fowler, P. Hedstr\"om, N. Jones, G. Kossinets, E. Lopez, A. Motter, M. A. Porter, S. Saavedra, M. Salganik, G. Szabo, and B. Uzzi for their comments on earlier versions of the manuscript, and to the workshop participants at the University of North Carolina at Chapell Hill, Northwestern, Northeastern, and Harvard University. J.-P. O. is supported by Wolfson College, University of Oxford, and the Fulbright Program. F. R.-T. acknowledges the support of EPSRC grant no. EP/E056997/1 and the European Commission FP7 FET project ICTeCollective (contract no. 238597).

The authors declare that they have no competing financial interests.

Correspondence and requests for materials should be addressed to JPO (email: Onnela@hcp.med.harvard.edu).
 
 
\clearpage{}
 
\section*{Supplementary Information}

\subsection*{S1: Background to fluctuation scaling}

Fluctuation scaling (FS) was introduced well to a wider physics audience in a recent article by Eisler,  Bartos, and Kertesz\cite{eisleragain}. In temporal fluctuation scaling (TFS), we start from a multitude of $M$ time series measured in the interval $[0,T]$ and assume that the constituents, i.e. the random variables making up the signal, are additive. The signals are divided into blocks of duration $\Delta t$, and for any block in the interval $[t, t+\Delta t)$ the signal can be decomposed as 
\begin{equation}
f_i^{\Delta t}(t)=\sum_{n=1}^{N_i^{\Delta t}(t)} V_{i,n}^{\Delta t}(t),
\label{eq:sum}
\end{equation}
where $N_i^{\Delta t}(t)$ is the number of constituents within the block, i.e. the number of random variables $V_{i,n}^{\Delta t}(t)$ to be summed together, of signal $i$ during $[t,t+\Delta t)$. We assume that $V_{i,n}^{\Delta t}(t) \ge 0$, so that the time average of $f_i^{\Delta t}$, denoted by $\ev{f_i^{\Delta t}}$, does not vanish. Is is defined as 
\begin{equation}
\ev{f_i^{\Delta t}}=\frac{1}{Q}\sum_{q=0}^{Q-1} f_i^{\Delta t}(q\Delta t) = \frac{1}{Q}\sum_{q=0}^{Q-1} \sum_{n=1}^{N_i^{\Delta t}(q\Delta t)} V_{i,n}^{\Delta t}(q\Delta t),
\label{eq:timeavg}
\end{equation}
where $Q=T/\Delta t$. For any $\Delta t$ the variance can be obtained as 
\begin{equation}
\sigma^2_i(\Delta t) = \ev{[f_i^{\Delta t}]^2}- \ev{f_i^{\Delta t}}^2.
\label{eq:var}
\end{equation}
This quantity characterizes the fluctuations of the activity of signal $i$ from block to block. When $f$ is positive and additive, it is often observed that the relationship between the standard deviation $\sigma_i(\Delta t)$ and the mean $\ev{f_i^{\Delta t}}$ is given by a power law
\begin{equation}
\sigma_i(\Delta t) \propto \ev{f_i^{}}^{\alpha_\mathrm{T}},
\label{eq:scaling}
\end{equation}
where one varies $i$ keeping $\Delta t$ fixed. Note that the value of $\Delta t$ does not affect the scaling as it can be absorbed in the proportionality constant. The exponent $\alpha_\mathrm{T}$ is in the range $[1/2,1]$ and the subscript $T$ indicates that the statistical quantities are defined as temporal averages to distinguish them from ensemble fluctuation scaling\cite{eisleragain}. 

In the paper, we discuss a more system specific form of fluctuation scaling using spin variables $S_{i,n}(t)$ as  constituent variables. Further, instead of having access to signals in continuous time, we consider, as a starting point, data sampled at discrete time intervals such that two consecutive time points $t$ and $t+1$ are separated by $\delta t$ in physical time. The corresponding events in real physical time may have an arbitrary time resolution but, due to finite temporal sampling resolution, all events within one block may be considered concurrent.

\subsection*{S2: Example of fluctuation scaling}

Let us consider a set of state or spin variables $S_{i,j}(t) \in \{-1, 0, 1\}$, one for each application $i$ of every user. Here $S_{i,j}(t) = {1}$ corresponds to user $n$ adopting application $i$ at time $t$, $S_{i,j}(t) = {0}$ corresponds to there being no activity from user $j$ regarding application $i$ at time $t$, and $S_{i,j}(t) = {-1}$ corresponds to user $j$ dropping application $i$ at time $t$. The FS exponent $\alpha$ can be interpreted in terms of correlations between the constituent variables, in this case the spin variables $S_{i,j}(t)$. This leads to two limiting cases. If the constituent variables are uncorrelated, one obtains square-root scaling with $\alpha=1/2$, whereas if the constituent variables are fully correlated, one obtains a linear scaling with $\alpha=1$.

Two simple examples will illustrate this interpretation.  Consider a variable $S_{i,j}(t)$ with the mean and variance of given by $\langle S_{i} \rangle$ and $\Sigma_{S_{i}}^{2}$, respectively. If the random variables $S_{i,j}(t)$ are independent and identically distributed for all $n$ and $t$, we obtain by the linearity of the expectation operator $E[\cdot]$, taken over time, that 
\begin{equation}
\mu_{i} =  E \left [f_{i}(t) \right ] = E \left [\sum_{j=1}^{N}S_{i,j}(t) \right ] = N E \left [S_{i,j}(t) \right ] = N \langle S_{i} \rangle
\end{equation}
The variance is given by
\begin{equation}
\sigma^2_{i} = \var \left [f_{i}(t) \right ] = \var \left [ \sum_{j=1}^{N}S_{i,j}(t) \right ] = N\var \left [S_{i,j}(t) \right ] = N\Sigma_{S_{i}}^{2}
\end{equation}
since the variance of the sum of uncorrelated random variables (as follows from their independence) is the sum of their variances. Combining the expression for the mean and the variance gives $\sigma^2_{i} = (\Sigma_{S_{i}}^{2}/\langle S_{i} \rangle) \mu_{i}$ so that $\alpha=1/2$. The exponent $\alpha = 1/2$ is then a consequence of the central limit theorem and is reminiscent of the $1/\sqrt{N}$ fluctuations of extensive quantities, such as energy, in equilibrium statistical mechanics\cite{eisleragain}. On the other hand, if the random variables $S_{i,j}(t)$ are completely correlated, such that $S_{i,1}(t) = \cdots = S_{i,N}(t)$, we can write $\sum_{j=1}^{N}S_{i,j}(t) = NS_{i,1}(t)$ which, as before, gives
\begin{equation}
\mu_{i} = N E \left [ S_{i,1}(t) \right ] = N \langle S_{i} \rangle
\end{equation}
but now 
\begin{equation}
\sigma^2_{i} = \var \left [ N S_{i,1}(t) \right ] = N^{2}\var \left [ S_{i,1}(t) \right ] = N^{2}\Sigma_{S_{i}}^{2},
\end{equation}
resulting in $\sigma_{i} = (\Sigma_{S_{i}}^{2}/\langle S_{i} \rangle) \mu_{i}$ so that $\alpha=1$. One way to produce  $\alpha=1$ is by a global driving force that imposes strong fluctuations that dominate over the local dynamics of the system\cite{eisleragain}.

\subsection*{S3: Stationarity of time series}

The fact that for most applications $n_{i}(t)$ is an increasing function of time suggests that the system is not stationary and, consequently, violates the assumption on stationarity. The question then becomes whether the system is sufficiently close to stationarity so that the fluctuation scaling exponents can be interpreted in terms of correlations among the constituent variables. We can write
\begin{equation}
\mu_{i} \equiv \ev{f_i(t)}=
\frac{1}{T_{i}}\sum_{t=1}^{T_{i}} f_i(t) = 
\frac{1}{T_{i}}\sum_{t=1}^{T_{i}} \left [ n_{i}(t) - n_{i}(t-1) \right ] =
\frac{1}{T_{i}}\sum_{t=1}^{T_{i}} \sum_{j=1}^{N} S_{i,j}(t), 
\label{eq:f}
\end{equation}
where the latter sum is taken over all $N$ Facebook users and we have used $\sum_{n=1}^{N} S_{i,j}(t) = n_{i}(t) - n_{i}(t-1)$. Let us now assume that only irreversible $S_{i,j}(t) = 0 \to S_{i,j}(t+1)=1$ changes are possible. The validity of this assumption has mostly to do with the choice of the investigated time period. Facebook applications had just recently been introduced, there was less choice of and less competition between applications and, hence, dropping of applications was conceivably rather rare. Quantifying the extent of uninstallation of applications would, however, require access to the micro level data.

Instead of letting the sum indexed by $n$ in the equation run over the entire system (over all users), we construct a restricted sum consisting of those users only who have not adopted application $i$ by the previous time step. This yields 
\begin{equation}
\mu_{i} = \frac{1}{T_{i}}\sum_{t=1}^{T_{i}} \sum_{j=1}^{N} S_{i,j}(t) 
\approx  \frac{1}{T_{i}}\sum_{t=1}^{T_{i}} \sum_{k=1}^{N-n_{i}(t)} S_{i,j_{k}}(t),
\label{eq:fa}
\end{equation}
where the subset of indices $j_{1}, j_{2}, \ldots, j_{N-n_{i}(t)} \in \{1, 2, \ldots, N\}$ such that $S_{i,j_{k}}(t-1)=0$.

The non-stationarity of $f_{i}(t)$ is reflected in the fact that the number of terms in the above sum,  $N-n_{i}(t)$, depends on (typically decreases with) time. While this is true for almost every application, it may be a problem only for the highly popular applications, i.e. in the high density regime. Let us impose the stringent condition that the system is within the low density regime, corresponding to the set of applications for which $f_{i}(t)$ are sufficiently close to stationarity, when at most 1\% of users have the application. Within this regime, the number of terms in the last sum of Eq.~\ref{eq:fa} is always between $0.99N$ and $N$ and, consequently, it decreases only marginally and the time series can be taken to be sufficiently stationary.

To see how far the low density regime extends, we set $N - n^{*} = 0.99N$, giving an upper limit $n^{*} = N/100$. The number of users at the end of the time period is $n_{i}(T) = n_{i}(0) + \mu_{i}T \approx \mu_{i}T$, the approximation being rather good in the low density regime, and we can assume that the approximate stationarity holds throughout the time horizon for applications with $n_{i}(T) \le n^{*}$, and setting $n^{*} = \mu^{*}T$ defines the low density regime as $0 < \mu < \mu^{*}$ with $\mu^{*} = N / (100T)$. The stationarity can be expected to break down for applications with $\mu_{i} > \mu^{*} \approx 414$ so that $\log(\mu^{*}) \approx 2.6$. This means that, even under this relatively strict interpretation of stationarity, $97.8\%$ of the time series are stationary. This also means that the scaling in Fig.~2C holds for over two orders magnitude \emph{above} the cross-over point $\mu_{x}$. We conclude that the system is sufficiently stationary so that the fluctuation scaling exponents for temporal fluctuations may be interpreted in terms of correlations between the constituent variables.

We can also relax the assumption about having only irreversible $S_{i,j}(t) = 0 \to S_{i,j}(t+1)=1$ changes. Let $S_{i,j}(t) = {1}$ correspond to user $j$ adopting application $i$ at time $t$, $S_{i,j}(t) = {0}$ corresponds to there being no activity from user $j$ regarding application $i$ at time $t$, and $S_{i,j}(t) = {-1}$ corresponds to user $j$ dropping application $i$ at time $t$. Allowing $S_{i,j}(t) = {-1}$ means that the value of $\ev{f_i}$ may vanish or become negative. Of the $M=2705$ applications analysed, 2562 have positive $\mu_{i} > 0$, 5 have $\mu_{i} = 0$, and for 138 applications $\mu_{i}  < 0$. Combining these numbers, we can see that $95\%$ of the temporal averages $\mu_{i} $ are, in fact, positive and, consequently, non-negativity does not pose a problem. 

\subsection*{S4: Breakpoint analysis for linear regression}

Consider the linear regression model

\begin{equation}
y_{i} = x_{i}^T \beta_{i} + u_{i}, \,\, i=1,\ldots,n, 
\end{equation}
where $y_{i}$ is observation $i$ of the dependent variable, $x_{i}$ is a $k \times 1$ vector of regressors with the first component set equal to unity, and $\beta_{i}$ is a $k \times 1$ vector of regression coefficients that may vary over time. The null hypothesis is that the regression coefficients remain constant
\begin{equation}
H_{0}: \, \beta_{i} = \beta_{0}, \,\, i = 1, \ldots, n
\end{equation}
against the alternative hypothesis $H_{1}$ that at least one of the coefficients changes. In general, if there are $m$ breakpoints, the regression coefficients are constant within the resulting $m+1$ segments. The model can be rewritten to incorporate the breakpoints as
\begin{equation}
y_{i} = x_{i}^T \beta_{j} + u_{i}, \,\, i=i_{j-1} + 1, \ldots, i_{j}, \,\, j=1,\ldots,m+1,
\end{equation}
where $\{i_{1}, \ldots, i_{m} \}$ are the set of breakpoints and $j$ is the segment index. Conventionally $i_{0} = 0$ and $i_{m+1} = n$. Breakpoints are typically not given exogenously but need to be estimated from the data. Finding breakpoints in data is also known as testing for structural change in data, and the are two frameworks for doing that: $F$-statistics and generalized fluctuation tests\cite{zeileis}. Here we follow the $F$-statistics test that can be used to test against a single breakpoint, corresponding to the case with $m=1$ in the above framework, at an unknown observation $i_{1}$ with segment $j=1$ covering observations $i=1, \ldots, i_{1}$ and segment $j=2$ covering observations $i=i_{1}+1, \ldots, n$. To identify the breakpoint $i_{1}$, we compute a sequence of $F$-statistics for a change at observation $i$ given by
\begin{equation}
F_{i} = \frac{\hat{u}^T \hat{u} - \hat{u}(i)^T \hat{u}(i)}{\hat{u}(i)^{T} \hat{u}(i) / (n-2k)}, 
\end{equation}
where $\hat{u}$ are the ordinary least squares residuals from the unsegmented (no breakpoint) model and $\hat{u}(i)$ are the ordinary least squares residuals from a segmented model with a breakpoint at observation $i$ and the regression is carried out separately for each segment\cite{zeileis}.

From the above definition it is clear that $F_{i}$ is proportional to the residuals of the unsegmented model, $\hat{u}^T \hat{u}$, and inversely proportional to the residuals of the segmented model, $\hat{u}(i) \hat{u}(i)^{T}$. To ensure that each regression model can be estimated with a sufficient number of data points, we need to introduce a trimming parameter $h$ such that we compute $F_{i}$ for a subset of $i=h, h+1, \ldots, n-h$ observations. In practice, we can compute $F_{i}$ for all $i=1, \ldots, n$ and simply ignore the resulting values of $F_{i}$ for very small and very large values of $i$, where a suitable value of $h$ is chosen by the practitioner. The null hypothesis $H_{0}$ is rejected if the maximum value of $F$ is ``large''\cite{zeileis}. What precisely it means for $F$ to be large depends on the context. In any case, what matters is the relative height and narrowness of the maximum value of $F$ with respect to all the other values; A peak that is high and narrow is stronger evidence of a structural change in data than a peak that is low and wide.

\begin{figure}
\includegraphics[width=1.0\linewidth]{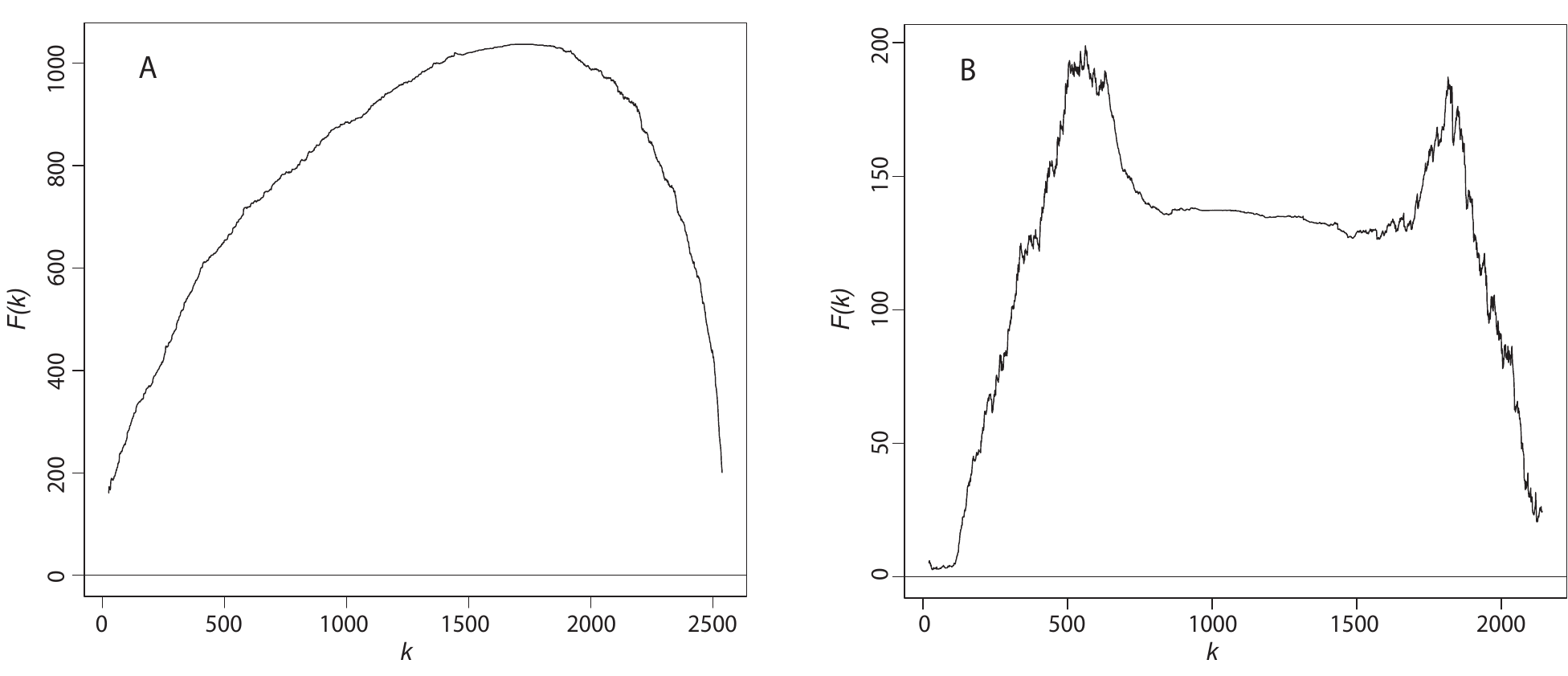}
\caption{$F$-statistics for breakpoint analysis. 
\textbf{(A)} The $F$-statistic is smooth and well-behaved for the empirical data and reaches a maximum of $F_{(k)} \approx 1035$ for observation $k=1759$. This maximum corresponds to a breakpoint at $\log(\mu_{(1759)}) \approx 0.36$ and it separates the data into two regimes characterized by exponents $\alpha_{I} \approx 0.55$ and $\alpha_{C} \approx 0.85$ to the left and right of the point, respectively. 
\textbf{(B)} The $F$-statistic for the synthetic data is very rugged and the resulting maximum is in practice degenerate. This irregular behavior of the $F$-statistic violates the underlying assumption of having a well-defined maximum and, consequently, does not provide sufficient statistical evidence for introducing a breakpoint in the data.
\label{fig:ftests}}
\end{figure}

The results are shown in Figure \ref{fig:ftests}. The data have been sorted in ascending order based on the $x$-variable such that $\mu_{(1)} \le \mu_{(2)} \le \cdots \le \mu_{(M)}$. In the case of the empirical data, the $F$-statistic behaves smoothly and develops a clear maximum. This is strong evidence of there being a structural change in the data such that the two regimes to the left and right of the breakpoint are governed by different exponents, $\alpha_{I} \approx 0.55$ and $\alpha_{C} \approx 0.85$, respectively.

The behavior of the $F$-statistic for the synthetic data, however, is qualitatively very different. Instead of a smooth, single maximum, the error landscape is more rugged, and the maximum appears to be degenerate. Strictly speaking, there is a single maximum at $F_{(k)} \approx 186$ for observation $k=562$, corresponding to $\log(\mu_{(562)}) \approx -0.38$, but there is also a secondary maximum for $k \approx 1800$. The lack of a clearly defined maximum suggests that there is no sufficient statistical evidence to introduce a breakpoint in the data. Note that the above framework does not allow introducing multiple breakpoints. While this could be done in principle by adding more degrees of freedoms (more parameters), it becomes exceedingly difficult to justify them, especially if the differences in the slopes are very small. To demonstrate this, consider accepting the view that there is, in fact, a legitimate breakpoint at $k=562$ in the synthetic data. This results in two exponents  $\alpha \approx 0.84$ and $\alpha \approx 0.87$, which are so close to one another that it is difficult to justify theoretically their slightly different values. We conclude, given these considerations, that the behavior of the synthetic data is governed by just a single exponent $\alpha_{S} \approx 0.84$. 

\subsection*{S5: Supporting data analysis}

\begin{figure}
\begin{center}
\includegraphics[width=1\linewidth]{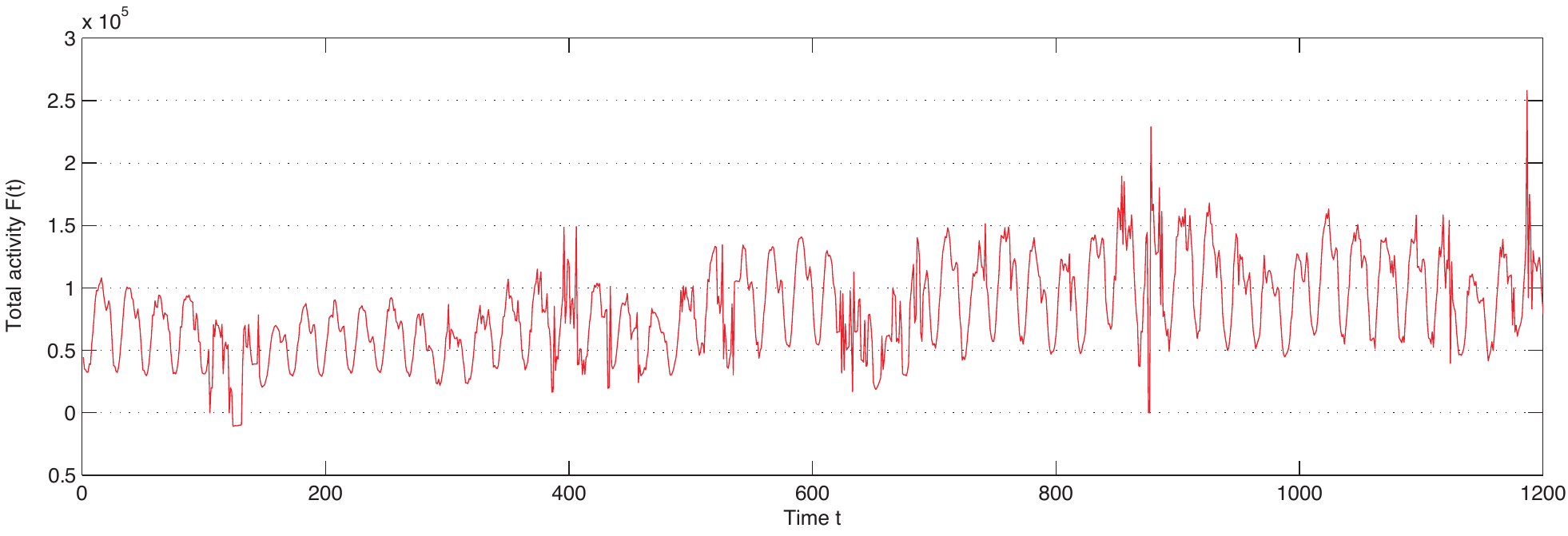}
\caption{Total activity $F(t)$ as a function of time $t$, where a unit of time is one observation, corresponding to calendar time from June 25, 2007 to August 14, 2007.
\label{fig:newsi1}}
\end{center}
\end{figure}

Let us define the \emph{total activity} as $F(t) = \sum_{i} f_{i}(t)$, where the sum runs over all applications that are in existence at time $t$. The total activity $F(t)$, which is not to be confused with the $F$-statistic in Section S4, corresponds to the total number of applications installed in the one-hour interval between $t$ and $t-1$. We show $F(t)$ in Fig.~\ref{fig:newsi1}, where the daily 24-hour period of activity is clearly visible.

\begin{figure}
\begin{center}
\includegraphics[width=0.9\linewidth]{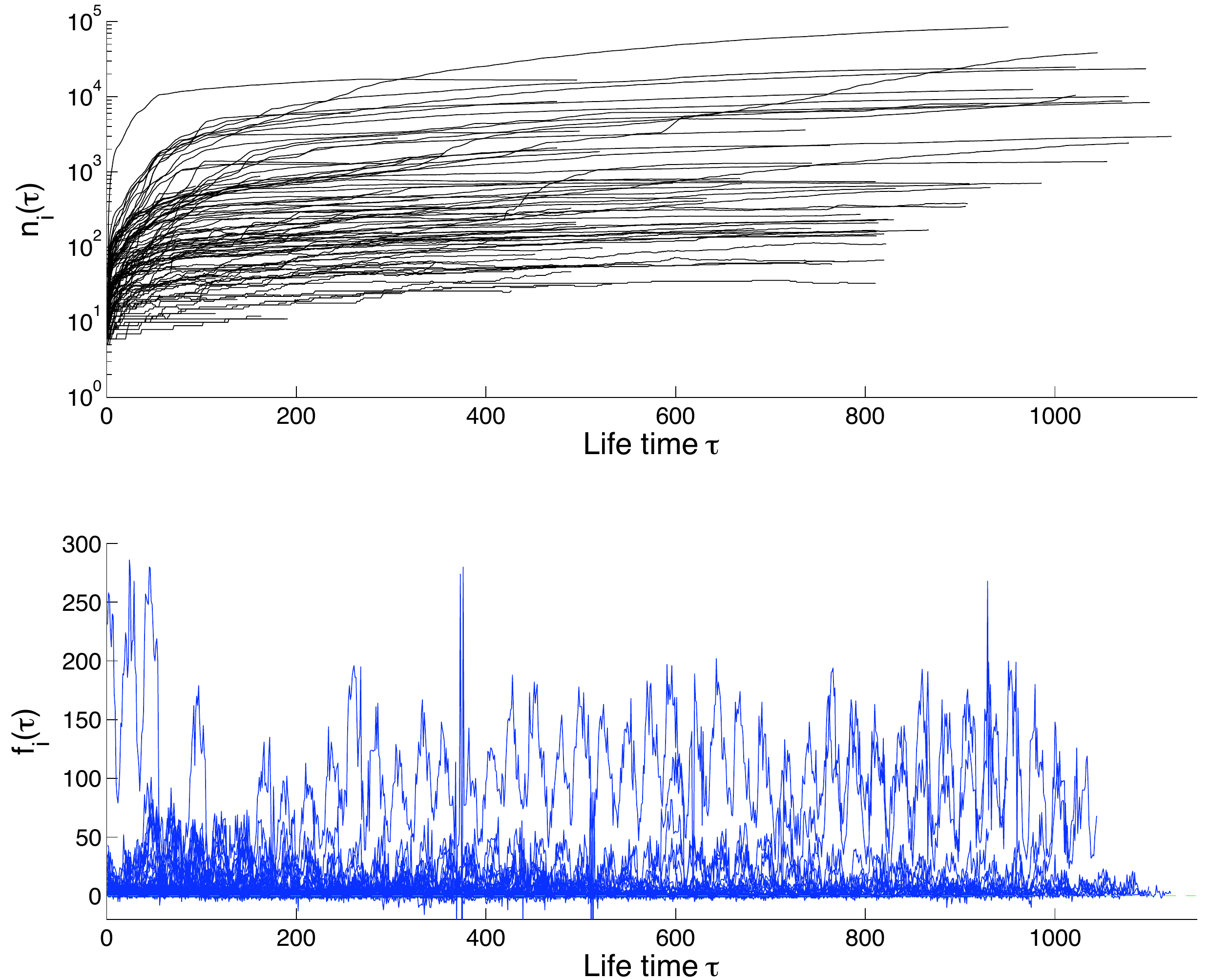}
\caption{Time-shifted aggregate numbers of application installations $n_{i}(t_{i} + \tau)$ as a function of application lifetime $\tau$ (top) and the related time-shifted activity values $g_{i}(\tau)$ (bottom). For purposes of visualization, in both plots are included a subset of 100 applications.
\label{fig:newsi2}}
\end{center}
\end{figure}

\begin{figure}
\begin{center}
\includegraphics[width=0.8\linewidth]{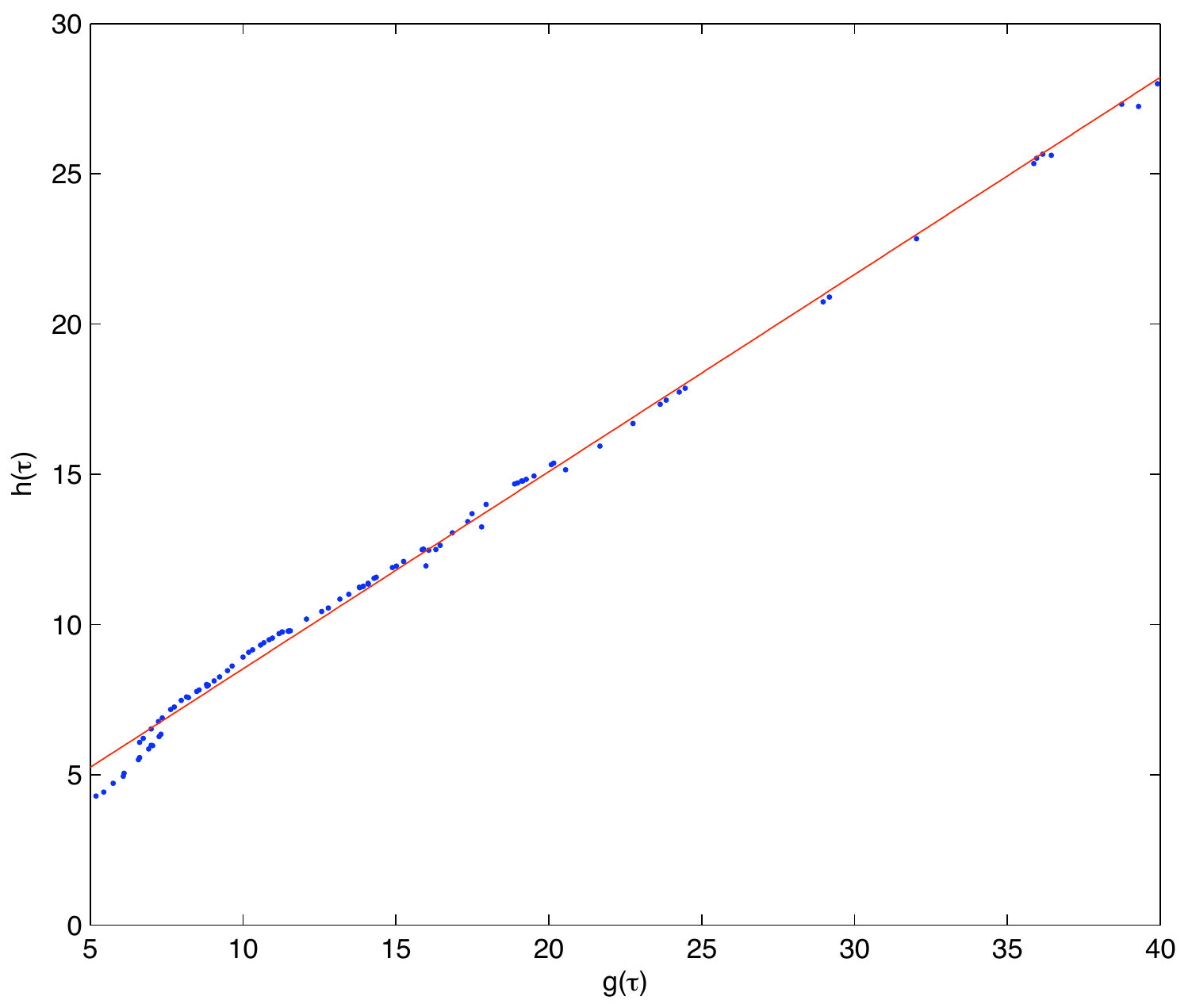}
\caption{Ensemble average of the standard deviation of the time-shifted activity, $h(\tau)$, has a fixed dependence on the ensemble average of the time-shifted activity, $g(\tau)$, throughout the lifetime of applications. The different points correspond to different values of $\tau \in {50, 60, 70, \ldots, 1000}$ such that increasing the value of $\tau$ leads to increasing values of $g(\tau)$. The values of $\tau$ start at 50 since we required that for each application there should be at least 50 points in the time series in order to estimate its first and second moments sufficiently accurately.
\label{fig:newsi3}}
\end{center}
\end{figure}

\begin{figure}
\begin{center}
\includegraphics[width=1.0\linewidth]{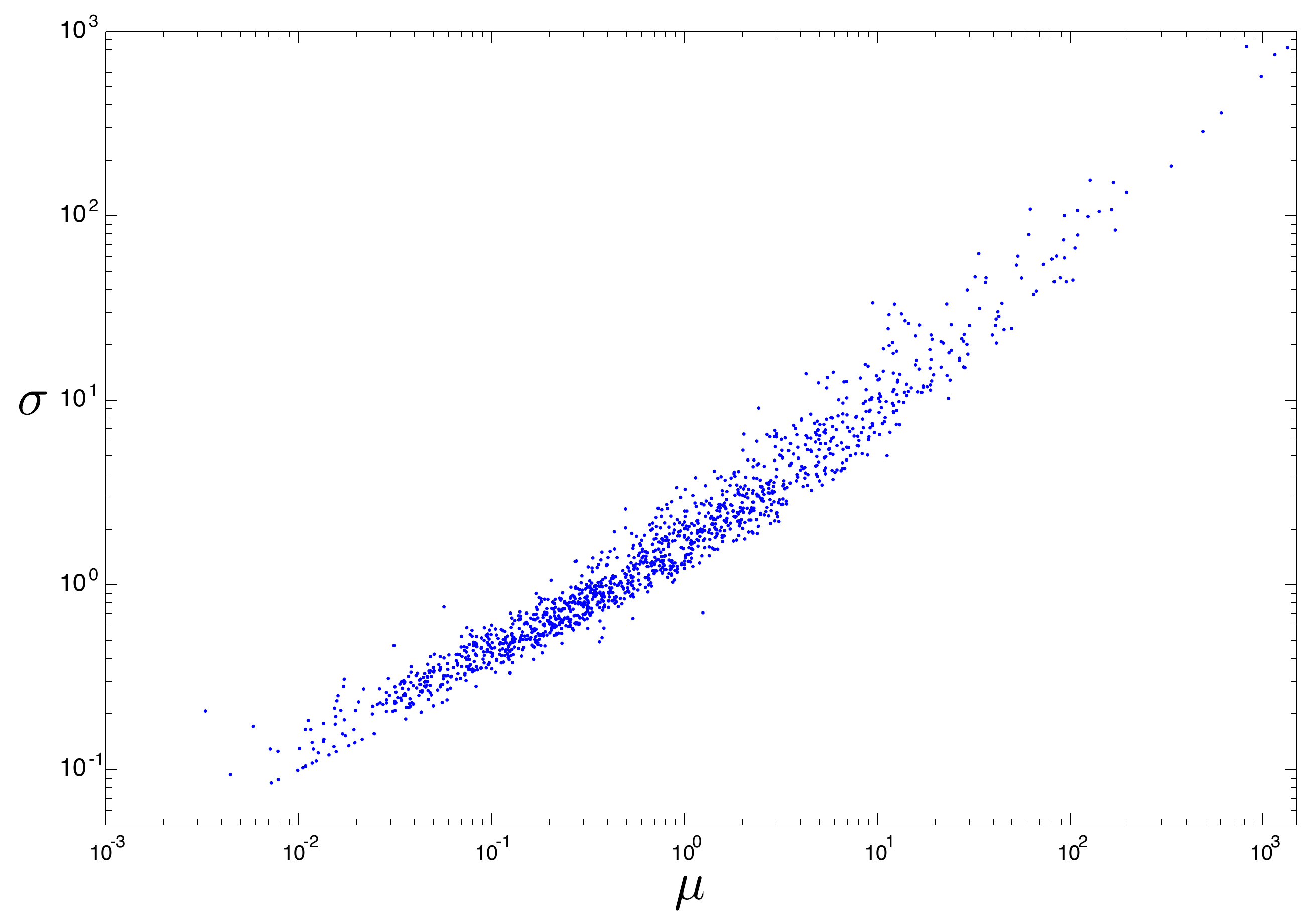}
\caption{Fluctuation scaling plot for activity $f_{i}$ using only those applications $i$ that were introduced during the studied time period and for which we had at least 50 time steps worth data.
\label{fig:newsi4}}
\end{center}
\end{figure}

It is possible that, for a given application, the mean and standard deviation of activity $f_{i}$ result from the application being at a certain stage of its lifetime. Consequently, given that we have a mixture of old and new applications, if the scaling of standard deviation of $f_{i}$ with the mean of $f_{i}$ were dependent on the age of the application, this could in principle contribute to the cross-over reported in the main text. To test this hypothesis, we define the time shifted activity for application $i$ as $g_{i}(\tau) = f_{i}(t_{i} + \tau)$ with $\tau \ge 0$, where $t_{i}$ is the (approximate) introduction time of application $i$. The time-shifted aggregated numbers $n_{i}(t_{i} +\tau)$ are shown in the upper panel of Fig.~\ref{fig:newsi2}, and the time-shifted activities $g_{i}(\tau)$ are in the lower panel. We can now compute the mean and standard deviation of the time-shifted activity $g_{i}(\tau)$ by truncating the time series at $\tau$, i.e. by taking the first $\tau$ points of the time series.  We define an ensemble average of the time-shifted activities taken over all $N(\tau)$ applications that have a lifetime of at least $\tau$ as
\begin{equation}
g(\tau) = \frac{1}{N(\tau)} \sum_{i=1}^{N(\tau)} \frac{1}{\tau} \sum_{t=1}^{\tau} g_{i}(t).
\end{equation}
Similarly, we can define the ensemble average of the standard deviation of the time-shifted activities as 
\begin{equation}
h(\tau) = \frac{1}{N(\tau)} \sum_{i=1}^{N(\tau)} \left [ \frac{1}{\tau-1} \sum_{t=1}^{\tau} \left (g_{i}(t) - \langle g_{i}(\tau) \rangle \right )^{2} \right ]^{1/2},
\end{equation}
where $\langle g_{i}(\tau) \rangle = (1/\tau) \sum_{t=1}^{\tau} g_{i}(t)$. We plot the standard deviation $h(\tau)$ versus the mean $g(\tau)$ for a number of different truncation points $\tau \in {50, 60, \ldots, 1000}$ in Fig.~\ref{fig:newsi3}. A linear fit describes their dependence very well, and demonstrates that the relationship between the mean and the standard deviation for the ensemble of applications does not depend on the age of the application, i.e. the stage the application in its lifetime. The fact that the dependence of $h(\tau)$ on $g(\tau)$ holds throughout the measured lifetime of applications demonstrates that the cross-over in the fluctuation scaling plot in the main text cannot be explained by having a mixture of applications that are at different stages of their lifetime. Finally, we repeat the fluctuation scaling plot in Fig.~\ref{fig:newsi4}, this time using only applications that have lifetimes $50 \le \tau_{i} \le T$ such that they were introduced during during the first $T-50$ time steps, corresponding to $t_{i} \in [0, T-50]$, such that for each application we have at least 50 points for estimating the first and second moments. The result is essentially identical to the one presented in the main text. In particular, the high $\mu$ applications are still present, as is the cross-over (fits not shown). This demonstrates explicitly that the high $\mu$ regime is not simply produced by applications that have a large number of installations for $t < 0$, i.e. before the start of data collection.

\bibliographystyle{Nature}

\begin{thebibliography}{23}
\bibitem{jpnote} Facebook has discontinued this practice after the data were collected.
\bibitem{footnote1} We estimate the mean and standard deviation of the entire activity time series using $\langle f_{i} \rangle \equiv \mu_{i} = \frac{1}{T_{i}}\sum_{t=1}^{T_{i}} f_{i}(t)$ and $\sigma_{i} = \left( \frac{1}{T_{i}-1} \sum_{t=1}^{T_{i}} [f_{i}(t) - \langle f_{i} \rangle]^{2} \right)^{1/2}$, where $T_{i}$ is the application specific time series length reflecting the fact that different applications were introduced at different times.
\end{thebibliography}

\bigskip
\hrule

\section*{Author Contributions}

JPO \& FRT formulated the underlying ideas; JPO performed the data analysis and simulations; JPO \& FRT wrote the manuscript.

\end{document}